\documentclass[12pt]{iopart}
\usepackage{iopams}
\usepackage{amssymb}
\usepackage{graphics}
\usepackage{graphicx}

\begin{document}
\title[Ground state cooling of a nanomechanical resonator via a Cooper pair box qubit]
{Ground state cooling of a nanomechanical resonator via a Cooper pair box qubit}

\author{Konstanze Jaehne, Klemens Hammerer and Margareta Wallquist}
\address{Institute for Theoretical Physics, University of Innsbruck, and \\
Institute for Quantum Optics and Quantum Information of the Austrian Academy of Sciences, 6020 Innsbruck, Austria.}
\ead{konstanze.jaehne@uibk.ac.at}
\begin{abstract}
In this paper we present a scheme for ground state cooling of a flexural mode of a nanomechanical beam incorporated in a loop-shaped Cooper-pair box (CPB) circuit. Via the Lorentz force coupling of the beam motion to circulating CPB-circuit currents, energy is transferred to the CPB qubit which acts as a dissipative two-level system. The cooling process is driven by a detuned gate-voltage drive acting on the CPB. We analyze the cooling force spectrum and present analytical expressions for the cooling rate and final occupation number for a wide parameter regime. In particular, we find that cooling is optimized in a strong drive regime, and we present the necessary conditions for ground-state cooling.
\end{abstract}
\maketitle


\section{Introduction}

Cooling of a macroscopic mechanical resonator to its quantum mechanical ground state has
received immense experimental
\cite{Arcizet2006Nature,Arcizet2006PRL,Gigan2006,HoehbergerMetzger2004,Schliesser2006,Kleckner2006}
and theoretical
\cite{WilsonRae2004,Marquardt2007,Genes2008,WilsonRae2007,Martin2004,Zhang2005,Wang2007}
attention. The quest for quantum limited control of micro- or nanomechanical oscillators
is motivated by fundamental interest \cite{Schwab2005,Marshall2003} as well as possible
applications in quantum information
\cite{Mancini2003,Vitali2007,Rabl2004,Geller2005,Armour2002,Tian2005,Siewert2005} and
precision measurement \cite{Arcizet2006Nature,Arcizet2006PRL,LaHaye2004}. Despite these
promising and significant experimental and theoretical efforts, to date occupation
numbers below 1 have not been achieved.

One can distinguish roughly three routes towards cooling of a nanomechanical resonator
down to the quantum regime, which have been proposed or partly demonstrated: passive
optical cooling
\cite{Gigan2006,HoehbergerMetzger2004,Schliesser2006,Marquardt2007,WilsonRae2007}, active
optical feedback cooling \cite{Arcizet2006Nature,Arcizet2006PRL,Kleckner2006,Cohadon1999}
and cooling via coupling to a different heavily damped solid state system, such as a
superconducting single electron transistor \cite{Naik2006,Blencowe2005,Clerk2005}, a
quantum dot \cite{WilsonRae2004}, a Cooper-pair box \cite{Martin2004,Zhang2005} or
 a flux qubit \cite{Wang2007}.

For passive optical cooling it was shown theoretically in
\cite{WilsonRae2004,Marquardt2007,Genes2008} that cooling to the quantum mechanical
ground state is possible. Experimental work \cite{Schliesser2006} impressively
demonstrated this cooling method, but did not achieve the ground state due to
insufficient laser phase noise stabilization. 

For nanomechanical resonators coupled to superconducting circuits less is known about
ground state cooling, despite recent experimental achievements: It has been shown in
\cite{LaHaye2004,Naik2006,Knobel2003} that it is possible to integrate a nanoresonator
into small superconducting circuits and to excite only a single mode of the resonator
motion. Also, cooling effects have been demonstrated by coupling to single-electron
transistors \cite{Naik2006} and superconducting microwave cavities \cite{Teufel2008}. On
the theoretical side, \cite{Martin2004,Zhang2005} showed in a fully quantum mechanical
treatment that it is possible to reach the ground state for mechanical frequencies above
$100\,{\rm MHz}$ via capacitive coupling to a Cooper-pair box. For other circuit setups a
quantum treatment \cite{Blencowe2007,Clerk2005} predicts cooling, but final occupation
numbers beyond the ground state. Cooling effects were also predicted with semi-classical
approaches in \cite{Wang2007,Wineland2006}.

In this paper we perform a full quantum mechanical treatment proving that ground state
cooling is possible for nanomechanical resonators coupled via Lorentz force to a
loop-shaped Cooper-pair box qubit. The loop-shaped Cooper-pair box was originally
introduced to allow for qubit state read-out at an optimal working point \cite{Vion2002},
where the qubit eigenstates correspond to clockwise and anti-clockwise currents which can
be excited and measured. We investigate the conditions for ground-state cooling of the
flexural beam motion, using the decay $\Gamma$ of the excited Cooper-pair box qubit state
to dissipate the resonator energy. With carefully chosen qubit-drive detuning, energy
moves from the resonator motion to the qubit, which acts as an additional reservoir at a
lower effective temperature. We find that the final occupation number crucially depends
on the initial thermal occupation $N_{\rm m}$ of the resonator mode and the dimensionless
parameter $\beta= g^2/(\Gamma\gamma_{\rm m})$, with $\gamma_{\rm m}$ the intrinsic
dissipation rate of the resonator and $g$ the resonator-qubit coupling strength. Most efficient cooling is achieved for large $\beta$, which corresponds to the strong coupling regime of cavity QED. Our analysis shows that cooling is optimized in the strong-drive regime for the qubit, and ground state cooling is possible even for 10 MHz mechanical frequency.

Our quantum mechanical model is developed in close analogy to the description of laser
cooling of trapped ions \cite{Stenholm1986,Cirac1992}. The basic working principle there
is that a fast, dissipative dynamics of the internal degree of freedom couples weakly to
the motional degree of freedom and provides an effective dissipative evolution for the
latter. This is translated here to the solid state context, where the CPB takes over the
role of internal electronic levels and the mechanical resonator corresponds to the ion
motion in a harmonic trapping potential, as was done in \cite{WilsonRae2004,Martin2004}.
Readers not so familiar with the quantum optical techniques of master equations and
adiabatic elimination based on projector operator formalism will find many useful details
of calculations in the appendix to this paper.

This paper is organized as follows. In section \ref{SetupSection} we present the
nanoresonator beam and the CPB circuit, and derive the Lorentz force interaction term.
Section \ref{CoolingSection} is devoted to a quantum mechanical derivation of the cooling
equation, starting from a master equation of the full system. In section \ref{Discussion}
we discuss the cooling rate and the final occupation number of the resonator motion in
various parameter regimes, before we conclude in section \ref{Conclusion}.

\section{The setup\label{SetupSection}}

In this paper we study a nanomechanical resonator beam integrated into a loop-shaped
Cooper-pair box (CPB) circuit, with the goal to derive conditions for ground-state
cooling of a flexural mode of the nanoresonator via the Lorentz force. The CPB circuit here acts as a
dissipative two-level system (qubit) which removes energy during the cooling process. In
this section we present a model for the motion of the nanoresonator and describe the
CPB circuit with the integrated mechanical beam. The system Hamiltonian consists of three parts,
\begin{equation}
H = H_{\rm m} + H_{\rm q} + H_{\rm int}
\end{equation}
where $H_{\rm m}$ describes the dynamics of the nanoresonator, $H_{\rm q}$ represents
the driven CPB qubit and $H_{\rm int}$ describes the interaction between the motion of
the nanoresonator and the CPB qubit. The
coupling to the environment will be included using a master equation formalism.

\subsection{The nanomechanical resonator}

The nanomechanical resonator \cite{Cleland2003} considered here is a quasi-1D beam of
$\mu$-scale length that executes mechanical oscillations. In this setup we consider
a so-called doubly clamped beam - a beam fixed at both ends - which can undergo
longitudinal, torsional and flexural oscillations. We are mainly interested in flexural
oscillations. Assuming a homogeneous rod, the frequency for the $n$-th mode of flexural
oscillations is given by $\omega_n=\sqrt{E M_{\rm B}/(\rho A)}\beta_n^2 $
\cite{Cleland2003}, where $E$ is Young's modulus, $\rho$ the density and $A$ is the cross
section of the beam. $M_{\rm B}=h^3w/12$ is the bending moment for a beam of
rectangular cross section of width $w$ and height $h$, oscillating vertically.
$\beta_n L$ is a numerical factor which for the lowest mode is $\beta_1\simeq 4.73/L$,
where $L$ is the length \cite{Cleland2003}. In the following we restrict our analysis to
the lowest flexural mode, described by a harmonic oscillator
Hamiltonian $H_{\rm m}$,
\begin{eqnarray}
H_{\rm m}=\hbar\omega_{\rm m} a^\dag a,
\end{eqnarray}
with $a$ the annihilation operator satisfying bosonic commutation relations
$[a,a^\dag]=1$. With the beam integrated into a superconducting circuit, we assume an
aluminum construction with typical values $E=70$ GPa and $\rho=2.7 \,{\rm g}/{\rm cm}^3$.
Choosing a height $h=0.3$ nm and a length $L=1\mu$m, we get a typical frequency
$\omega_{\rm m} =$10 MHz. Note that the frequency does not depend on the width $w$, but
we have to choose $w\ll L$ since we treat the nanorod as a quasi-1D beam.

The dynamics of the nanoresonator mode is further governed by coupling to phonons in the
electrical circuit and by the residual coupling to all other modes. Assuming that the
lowest flexural mode is sufficiently well isolated, we model the effect of its environment as the interaction with a thermal bath,
which can be treated in terms of an effective master equation \cite{MathMethods} with the Liouvillian $\mathcal{L}_{\rm m,d} (\rho)$,
\begin{equation}
\mathcal{L}_{\rm m,d} (\rho) = \gamma_{\rm m}(N_{\rm m}+1) {\cal D}[a](\rho) +
\gamma_{\rm m} N_{\rm m} {\cal D}[a^\dag](\rho) \label{MasterEqNanorodDiss}
\end{equation}
where
\begin{equation}
{\cal D}[A](\rho)=2A\rho A^\dag-A^\dag A\rho-\rho A^\dag A\label{DecayOperator}.
\end{equation}
The decay rate $\gamma_{\rm m} =\omega_{\rm m}/Q =$ 0.1 kHz for a typical Q-value of
$10^5$. At cryogenic temperatures ($T=15$ mK) the steady state of the nanoresonator is a
thermal state corresponding to an occupation number $N_{\rm m} = [\exp(\hbar\omega_{\rm
m}/(k_{\rm B} T))-1]^{-1} \simeq$ 200.


\subsection{The Cooper pair box as a dissipative two level system}

\begin{figure}
\begin{centering}
\includegraphics[scale=0.8]{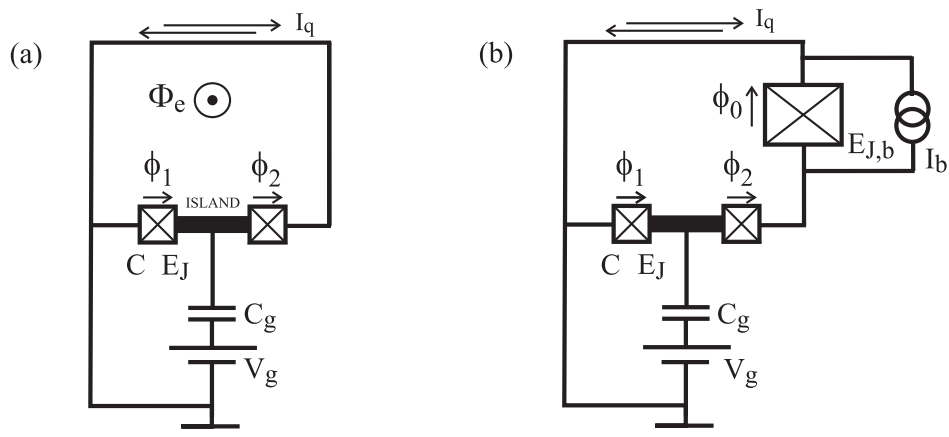}
\caption{Two ways of creating currents in a split CPB circuit. The CPB
consists of a superconducting island coupled by two Josephson junctions, each with Josephson energy $E_J$
and capacitance $C$, to the superconducting electrode. For a specific gate voltage $V_g$ the qubit eigenstates correspond to circulating
currents in the circuit. Currents are excited (a) by a magnetic flux $\Phi_e$ applied
through the loop or (b) by coupling the CPB to a large current-biased Josephson junction.
\label{SetupFigure}}
\end{centering}
\end{figure}

A Cooper Pair box (CPB) \cite{Lafarge1993,Shnirman1997,Makhlin2001} is a superconducting
island connected via one or two Josephson junctions (JJ) to a superconducting reservoir. The dynamics of
the CPB is governed by the charging energy $E_{\rm C}$, given by
$E_{\rm C}=(2e)^2/(2C_\Sigma)$ with $C_\Sigma$ the total capacitance of the island, and
the Josephson energy $E_{\rm J}$, which determines the tunneling of Cooper-pairs between
the island and the reservoir. The dynamics is restricted to controlled Cooper-pair
tunneling if the superconducting gap $\Delta_{\rm s}$ is larger than
the single electron charging energy, $\Delta_{\rm s} > E_{\rm C}/4$, and thermal
processes are weak, $k_{\rm B} T\ll E_{\rm C}$. Under these circumstances, it is
possible, as we will show in \ref{AppA}, to define a two-level system (qubit) of the
two lowest energy eigenstates of the island, $|g\rangle$ and $|e\rangle$. In this paper
we consider the split CPB \cite{Vion2002}, as shown in figure \ref{SetupFigure}, where
two (ideally) identical JJs, each with capacitance $C$ and Josephson
energy $E_{\rm J}$, couple the island to the reservoir. Transitions between the two
states are induced by a gate voltage microwave drive $\delta V_{\rm
g}\cos{\omega_{\rm dr}t}$,
applied via a gate capacitance $C_{\rm g}$ together with a static bias $V_{\rm g,0}$. As we have shown in \ref{AppA}, the truncated circuit
Hamiltonian reads, in a
frame rotating with the microwave drive $\omega_{\rm dr}$,
\begin{equation}
H_{\rm q} = -\frac{\hbar\Delta}{2}\sigma_z + \frac{\hbar\Omega}{2}\sigma_x, \label{HTLS}
\end{equation}
with the detuning $\Delta = \omega_{\rm dr} - 2 E_{\rm J}/\hbar$ and the drive strength
$\Omega= E_{\rm C} \delta n_{\rm g}/\hbar$. $H_{\rm q}$ is written in the eigenbasis at
the charge degeneracy point, defined by $C_{\rm g} V_{{\rm g},0}/(2e)=1/2$ and
$\delta n_{\rm g}=C_{\rm g} \delta V_{\rm g}/(2e)=0$. At this particular bias point the system is less sensitive to charge fluctuations
\cite{Vion2002,Ithier2005}, which would transform into fluctuations of the detuning
$\Delta$ and in principle could make the cooling procedure less efficient.
For a more realistic setup, one has to assume an asymmetry of a
few percent; we will discuss this issue in \ref{AppA}.

As already mentioned in the introduction, the loop-shape of the split CPB allows for
circulating currents in the circuit. In fact, the energy eigenstates at the charge
degeneracy point correspond to clockwise and anticlockwise loop currents
\cite{Wallquist2005}. In the CPB circuit as described so far there is no asymmetry in the
loop allowing the flow of a net current in either direction and therefore we have to add
elements which break the symmetry. One possibility is to
thread the circuit loop with an external magnetic flux $\Phi_{\rm e}$, as shown in figure
\ref{SetupFigure}(a). The magnetic field identifies a direction in the loop, and
consequently one current direction is energetically favorable, giving rise to net
currents in the circuit. The other option is to integrate a large current-biased
Josephson junction with Josephson energy $E_{\rm J,b}$ into the loop, as shown in figure \ref{SetupFigure}(b). As we show in
\ref{AppA}, the bias current $I_{\rm b}$ determines the phase across the large JJ, and
thus acts in a similar way as the magnetic flux $\Phi_{\rm e}$. For the purpose of this
paper, both ways to excite the loop currents are equivalent, and give rise to loop
currents $\pm I_{\rm q}$ ,
\begin{equation}
I_{\rm q} = {E_{\rm J} \over 2}\sin\left({\phi_0 \over 2} \right)
\label{Eq:Iq},
\end{equation}
with $\phi_0$ defined by $\Phi_{\rm e}=(\hbar/(2e))\phi_0$ and
$\sin\phi_0=(\hbar/(2e))I_{\rm b}/E_{\rm J,b}$, respectively. It is obvious from the expression (\ref{Eq:Iq})
that in the absence of magnetix flux or bias current, $\phi_0 =
0$ and the loop currents vanish. Apart from the excitation of loop currents, the energy
splitting is affected and thus the detuning $\Delta$ is modified,
\begin{equation}
\Delta \ \rightarrow \ \omega_{\rm dr} - \frac{2E_{\rm
J}}{\hbar}\cos\left(\frac{\phi_0}{2}\right).
\end{equation}
As a last remark, we would like to point out that this derivation holds also for $E_{\rm
C}\simeq E_{\rm J}$, where the qubit is even less sensitive to charge fluctuations
\cite{Vion2002,Ithier2005}. Due to the anharmonicity of the energy spectrum, it is
possible to truncate the Hilbert space to the two lowest energy eigenstates at the charge
degeneracy point $\delta n_{\rm g}=0$, and it can be shown that these states are
eigenstates also of the truncated loop-current operator
\cite{Wallquist2005,WallquistThesis}. However, more charge states will be involved in the
qubit eigenstates, and thus the parameters $\Delta$ and $\Omega$ are modified.

So far we have only described the coherent dynamics of the CPB. Due to its coupling to
the surrounding circuit leads and uncontrolled degrees of freedom within the circuit
\cite{Ithier2005}, the qubit is subject to dissipation. The dissipative dynamics of the qubit is determined by the
depolarization time $1/(2\Gamma)$, which describes the relaxation to the ground state,
and the dephasing time $1/(\Gamma+\Gamma_{\rm d})$ \cite{Ithier2005}. These two processes are included phenomenologically
\cite{Makhlin2001} into the qubit master equation,
\begin{equation}
\mathcal{L}_{\rm q,d}(\rho) = \Gamma {\cal D}[\sigma_-](\rho) +\frac{\Gamma_{\rm
d}}{4}{\cal D}[\sigma_z](\rho)\label{MasterEqQDiss}
\end{equation}
with ${\cal D}[\sigma_-](\rho)$ and ${\cal D}[\sigma_z](\rho)$ given by equation
(\ref{DecayOperator}).
For an energy splitting $E_{\rm J}/h \sim 20$ GHz \cite{Vion2002} at a cryostat
temperature $T$ of 15 mK, the thermal occupation number of the excited state is
negligibly small, $N_{\rm q} = [\exp(E_{\rm J}/(k_BT))-1]^{-1}\simeq 0$, and therefore
the heating term $N_{\rm q}{\cal D}[\sigma_+](\rho)$ can be neglected. Both $\Gamma$ and
$\Gamma_{\rm d}$ can vary dramatically between different samples and working points
\cite{Duty2004,Vion2002}. Since the relaxation is crucial for the cooling procedure, we
assume a typical value $\Gamma\sim1$ MHz \cite{Vion2002}. The derivations presented in
section \ref{CoolingSection}, \ref{AppB} and \ref{AppC} are valid for a finite
$\Gamma_{\rm d}$, but for transparency we restrict the discussion in section
\ref{Discussion} to the limit $\Gamma_{\rm d}/\Gamma \rightarrow 0$.

\subsection{Lorentz force interaction \label{Interaction}}
\begin{figure}[h]
     \begin{center}
         \includegraphics[scale=1]{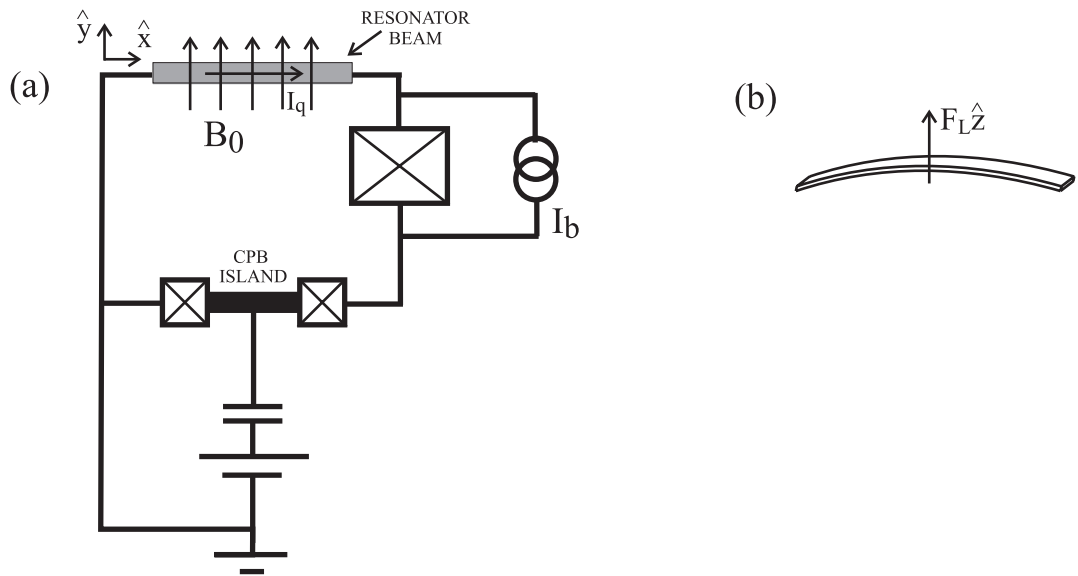}
     \end{center}
\caption{(a) Inclusion of the nanoresonator into the CPB circuit. The coupling between
the two systems will be provided by the application of a magnetic field along the
$\hat{y}$-axis in plane with the CPB, such that the current flowing in the circuit along the
$\hat{x}$-direction will couple to the motion of the oscillator via the Lorentz force,
which acts in the $\hat{z}$-direction as sketched in (b).\label{FullCircuitModel}}
\end{figure}

Now we include the nanoresonator beam into the loop of the CPB circuit as shown in figure
\ref{FullCircuitModel}(a).
As already mentioned earlier in this section, it is possible to excite circulating
currents in the circuit, and we assume a construction where the currents are also driven
through the resonator beam. A uniform magnetic field $B_0$ applied perpendicular to the
long axis of the beam, as shown in figure \ref{FullCircuitModel}(a), together with the
circulating qubit currents $I_{\rm q}$ (\ref{Eq:Iq}) creates a Lorentz force $F_{\rm L}$
which acts on the beam in the perpendicular direction $\hat{z}$
\cite{Cleland2003,LandauLifshitz,Cleland1996} as shown in figure
\ref{FullCircuitModel}(b),
\begin{equation}
\hat{F}_{\rm L} = B_0 \hat{I} L.
\end{equation}
Since the sign of the Lorentz force is qubit-state dependent through the qubit operator
$\hat{I}$ (\ref{Eq:Ioperator}), it can act as a cooling force on the displacement of the
beam in the $\hat{z}$-direction through the interaction term $H_{\rm int} = F_{\rm L}
\hat{z}$. Writing the displacement of the beam as $\hat{z} = a_0 (a + a^\dag)$, with
$a_0=\sqrt{\hbar/(2m\omega_{\rm m})}$ the amplitude of the zero-point motion, the
interaction Hamiltonian gets the form,
\begin{equation}
H_{\rm int} = \hbar g\sigma_z (a+a^\dag), \label{Eq:Hint}
\end{equation}
with the coupling constant $g$ given by,
\begin{equation}
g = \frac{B_0 I_{\rm q} L a_0}{\hbar} . \label{g0Def}
\end{equation}
Note that this interaction term differs from the ubiquitous Jaynes-Cummings interaction
\cite{WallsMilburn}, and thus the requirements for ground-state cooling will differ from
for example the case of ion cooling \cite{Stenholm1986}. In this paper we focus on the
case of constant $g$. Assuming $a_0\approx 10^{-13}$ m, $B_0 \approx 10$ mT and $I_{\rm
q} \approx 10$ nA, we achieve a typical coupling constant $g\approx 100$ kHz. In
principle it would also be possible to drive either the coupling magnetic field $B_0$ or
the bias current $I_{\rm b}$,  but this is beyond the scope of this paper. Note that the
coupling field $B_0$ does not create an additional flux in the circuit loop. However if
we choose to use the setup in figure \ref{SetupFigure}(a), the applied flux $\Phi_{\rm
e}$ in the $\hat{z}$-direction would create a Lorentz force in the perpendicular
direction $\hat{y}$ and could in principle lead to heating of the oscillation modes in
this direction. However this coupling is very weak, since the magnetic field required to
obtain $\Phi_{\rm e}$ is at least one order of magnitude lower than the coupling magnetic
field $B_0$. In addition, assuming a rectangular beam with $w \gg h$, the oscillation
frequency for the motion in the $\hat{y}$-direction would be large compared to
$\omega_{\rm m}$.

\section{Cooling\label{CoolingSection}}

In this section we derive an effective master equation for the relevant mode of the
resonator motion, and the corresponding cooling equation. Following the line of
\cite{Cirac1992}, our main assumption is $g\ll \Gamma$, which means that the qubit goes
to equilibrium on a time scale which is much shorter than the time scale for the
interaction between the qubit and the resonator. Thus the state of the qubit is hardly
affected by the interaction, and the qubit can be adiabatically eliminated. The second
important assumption is $\gamma_{\rm m} N_{\rm m}\ll\Gamma$. Due to the different time
scales for the dissipation of the resonator via the qubit and via the coupling to its own
environment, the two dissipation channels act separately.
We define the small parameters $\epsilon_1=g/\Gamma$ and $\epsilon_2=\gamma_{\rm m}N_{\rm
m}/\Gamma$, and thus treat the coupling between the qubit and the nanoresonator and also the
intrinsic dissipation of the resonator as perturbations. For the perturbation expansion of the
master equation, we write it on the following form,
\begin{eqnarray}
\dot{\rho}&=&\mathcal{L}_0^{\rm m}(\rho) +\mathcal{L}_0^{\rm q}(\rho) +\mathcal{L}_1 (\rho) +\mathcal{L}_2 (\rho) \label{MasterEq}
\end{eqnarray}
where the zeroth order terms describe the non-dissipative dynamics of the resonator,
\begin{equation}
\mathcal{L}_0^{\rm m}(\rho) = -\frac{i}{\hbar}[H_{\rm m},\rho]
\label{MasterEqL0m}
\end{equation}
and the dynamics of the qubit as described in section \ref{SetupSection},
\begin{equation}
\mathcal{L}_0^{\rm q}(\rho) = -\frac{i}{\hbar}[H_{\rm q},\rho]+\mathcal{L}_{\rm q,d}(\rho).
\label{MasterEqL0q}
\end{equation}
The first order term in $\epsilon_1$ describes the coupling between the qubit and the resonator,
\begin{equation}
\mathcal{L}_1 (\rho) = -\frac{i}{\hbar}[H_{\rm int},\rho]\label{MasterEqL1}
\end{equation}
whereas the term of first order in $\epsilon_2$ gives the coupling of the resonator to its environment as described in section \ref{SetupSection},
\begin{equation}
\mathcal{L}_2 (\rho)=\mathcal{L}_{\rm m,d} (\rho) \label{MasterEqL2}.
\end{equation}
We shift all details of calculations to \ref{AppB}, where we derive the reduced master equation for the motion of the resonator, and state here directly the result,
\begin{equation}
\dot{\bar{\rho}}_{\rm m} =-i\tilde{\omega}_{\rm m}[a^\dag a,\bar{\rho}_{\rm m}] +A_- {\cal D}[a](\bar{\rho}_{\rm m}) +
A_+ {\cal D}[a^\dag](\bar{\rho}_{\rm m})
\label{EQ:reducME}
\end{equation}
with $\bar{\rho}_{\rm m}={\rm tr}_{\rm q}\{\bar{\rho}\}$ given in a frame where the
resonator motion is coherently shifted as defined in (\ref{Eq:Shifta}), with the shift $\alpha$ given by (\ref{Eq:ShiftAlpha}). The effective frequency is
given by
\begin{equation}
\tilde{\omega}_{\rm m}=\omega_{\rm m}+g^2{\rm Im}\{S(\omega_{\rm m})+S(-\omega_{\rm
m})\},\label{Eq:omegameff}
\end{equation}
where the frequency shift is an order $\sim g^2/(\omega_{\rm m}\Gamma)\sim 10^{-3}$
smaller than the bare resonator frequency $\omega_{\rm m}$. The cooling and
heating rates are given by $A_-$ and $A_+$ respectively,
\begin{equation}
\fl A_-=2\left[g^2 {\rm Re}\{S(\omega_{\rm m})\}+\gamma_{\rm m}(N_{\rm m}+1)\right],
\qquad A_+=2\left[g^2 {\rm Re}\left\{S(-\omega_{\rm m})\right\}+\gamma_{\rm m} N_{\rm
m}\right]\label{Apm}.
\end{equation}
Here we defined the spectral function
\begin{equation}
S(\omega)=\int\limits_0^{\infty} d\tau e^{i\omega\tau}\langle
\delta\sigma_z(\tau)\delta\sigma_z\rangle_{\rm ss}, \label{ForceSpectrum}
\end{equation}
where $\delta\sigma_z=\sigma_z-\langle \sigma_z\rangle_{\rm ss}$ is an operator describing fluctuations of $\sigma_z$ about its mean steady state value $\langle \sigma_z\rangle_{\rm ss}$. The coupling to the CPB qubit has essentially two effects on the mechanical resonator, a negligibly small frequency shift and an additional dissipative channel, which can eventually exceed the natural dissipation due coupling to a phononic heat bath. If paramters are chosen in the right way, this can give rise to strong cooling. Note however that the master equation for the resonator motion (\ref{EQ:reducME}) is correct only up to $O(\epsilon_{1(2)}^2)$.

From the master equation we derive the corresponding equation for
the expectation value of the motional occupation number $\langle n \rangle$,
\begin{equation}
\frac{d}{dt}\langle n\rangle =-(A_--A_+)\langle n\rangle +A_+ . \label{CoolingEq}
\end{equation}
The cooling equation (\ref{CoolingEq}) gives us information about the total cooling rate $W$,
\begin{equation}
W = (A_--A_+) , \label{WRate}
\end{equation}
and, provided that we are in the cooling regime ($W>0$), the final occupation number is
given by
\begin{equation}
\langle n\rangle_{\rm f} = \frac{A_+}{W}. \label{nfEq}
\end{equation}
Equations (\ref{WRate}) and \ref{nfEq} are the main results of this paper.

%
%
\section{Discussion\label{Discussion}}

In the following, we will analyze the cooling rate (\ref{WRate}) and the final occupation
number (\ref{nfEq}), and define parameter regimes where cooling to the ground state is possible.
In \ref{AppC} we derive an expression for the spectrum $S(\omega)$ (\ref{SRApp}) which provides general expressions for the
cooling rate $W$ and for $\langle n\rangle_{\rm f}$ in terms of the system parameters
$\Delta, \Omega, \Gamma, \gamma_{\rm m}, \omega_{\rm m}, N_{\rm m}, g$. We use the full expression to generate plots for $\langle n\rangle_{\rm f}$
and $W$ for a realistic set of parameters which were motivated in section
\ref{SetupSection}. Furthermore we give analytic expressions in the limit of
weak qubit decay $\Gamma \ll \bar\Delta$, where we introduce
 the qubit energy splitting in the rotating
frame $\bar{\Delta}=\sqrt{\Omega^2 + \Delta^2}$ - see the single qubit Hamiltonian (\ref{HTLS}).

The cooling rate consists of two parts, as the coefficients $A_\pm$ (\ref{Apm}) show, where
one part is due to the intrinsic decay $\gamma_{\rm m}$ of the resonator and the other one is due to the coupling
$g$ to the dissipative qubit. The qubit serves as an additional bath for the
resonator, with an effective decay rate determined by the coupling strength $g$
and the real part of the force spectrum $S(\omega)$ (\ref{ForceSpectrum}). Thus it is the presence of the Lorentz force which makes it possible to cool the resonator motion below its thermal occupation.
In the limit of weak qubit decay rate, $\Gamma \ll \bar\Delta$, the real part of
the spectrum exhibits well-resolved peaks at $\pm \bar\Delta$ as we have shown in
\ref{AppC}.
\begin{figure}
\begin{centering}
\includegraphics[width=0.6\textwidth]{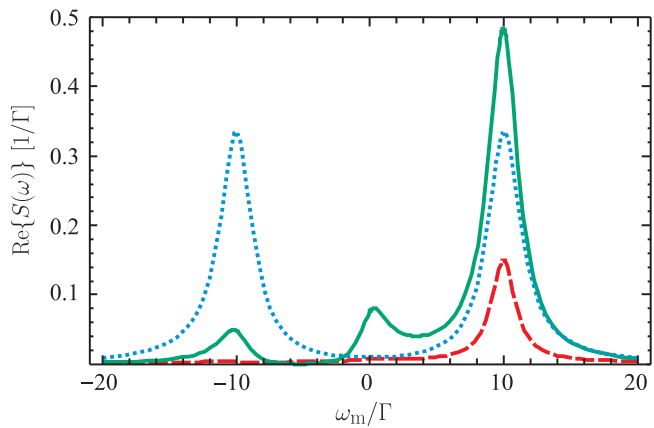}
\caption{Real part of the force spectrum (in units $1/\Gamma$) for the parameters:
$\omega_{\rm m}=10$ MHz, $\Gamma = 1$ MHz, $\bar{\Delta} = 10$ MHz, $\Omega =$ 4MHz
(dashed, red), $\Omega=8.5$ MHz (solid, green) and $\Omega=10$ MHz (dotted, blue). The
dashed curve shows the absence of the heating peak for weak drive, but still slow cooling
due to the small amplitude of the cooling peak. The solid line shows the optimal cooling,
where the cooling peak is high and the heating peak still small. For even stronger drive,
represented by the dotted line, the heating peak has increased so much that cooling and
heating balance each other and the net cooling effect is lost. \label{Fig:3Omega} }
\end{centering}
\end{figure}
To lowest order in $\Gamma/\bar\Delta$ the spectrum in the vicinity of the peaks can be
written,
\begin{equation}
{\rm Re}\left\{S(\omega \simeq \pm \bar\Delta)\right\} = {\Gamma_+ \alpha_\pm \over
\Gamma_+^2 + (\omega \mp \bar\Delta)^2}, \qquad \alpha_\pm = {\Omega^2 \over 2
\bar\Delta^2} \mp {\Delta \over \bar\Delta} {\Omega^2 \over 2 \Delta^2 + \Omega^2},
\label{Eq:alphas}
\end{equation}
with the effective decay rate $\Gamma_+$ presented in (\ref{GammaPlus}). Since only the
amplitude of the real part of the spectrum at the resonator frequency, ${\rm
Re}\left\{S(\pm\omega_{\rm m})\right\}$, enters the cooling equation (\ref{CoolingEq}),
it is intuitively clear that in this regime cooling is significant only for $\bar\Delta
= \omega_{\rm m}$. Note that ${\rm Re}\left\{S(\omega_{\rm m})\right\}$ is responsible
for cooling, whereas ${\rm Re}\left\{S(-\omega_{\rm m})\right\}$ contributes to heating.
When inserting equation (\ref{Eq:alphas}) into the cooling rate (\ref{WRate}), we observe
that a net cooling effect requires $\alpha_+
> \alpha_-$, i.e. negative detuning $\Delta < 0$, which will be assumed in the following.
We plot the real part of the spectrum as a function of the dimensionless frequency
$\omega/\Gamma$ in figure \ref{Fig:3Omega} for $\bar\Delta = \omega_{\rm m}$ and
different values for $\Omega$. The dashed curve for which we chose
$\Omega=0.4\,\omega_{\rm m}$ shows the absence of a heating peak for weak drive, but
still slow cooling due to the small amplitude of the cooling peak. The solid line
($\Omega=0.85 \omega_{\rm m}$) shows the optimal cooling, where the cooling peak is high
and the heating peak still small. For even stronger drive ($\Omega=\omega_{\rm m}$),
represented by the dotted line, the heating peak has increased so much that cooling and
heating balance each other and the net cooling effect is lost.

\begin{figure}[h]
     \begin{center}
         \includegraphics[scale=0.6]{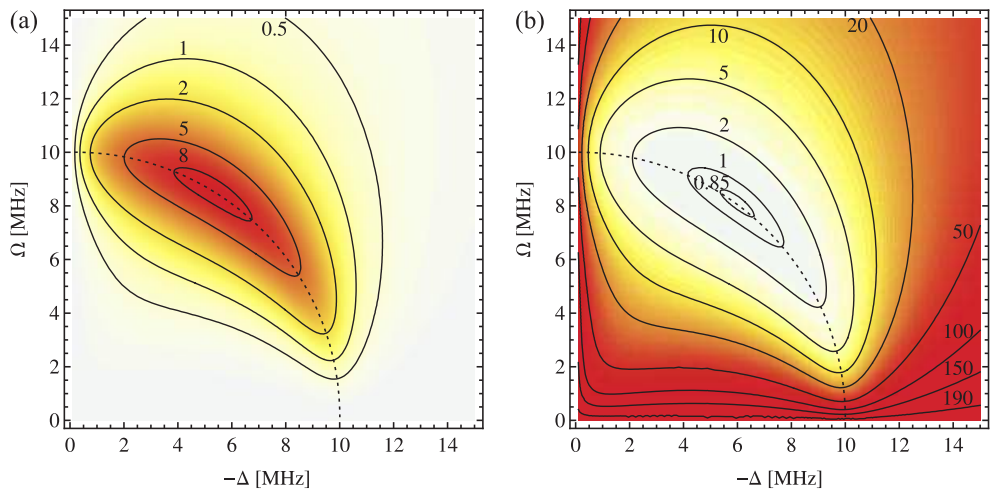}
     \end{center}
\caption{Contour plot of (a) the cooling rate (in units of kHz) and (b) the final occupation number as a
function of the detuning $\Delta$ and the drive strength $\Omega$ for the following
choice of parameters: $N_{\rm m}=195$ , $\gamma_{\rm m} =17$ Hz, $\Gamma=1$ MHz,
$\omega_{\rm m}=10$ MHz, $g=0.1$ MHz. The dotted line shows the special value
$\Omega^2+\Delta^2=\omega_{\rm m}^2$, where cooling is optimal. \label{PlotFigure1}}
\end{figure}

The total cooling rate as a function of detuning $\Delta$ and drive strength $\Omega$ is shown in figure \ref{PlotFigure1}(a).
We observe that cooling is obtained for a negative detuning ($\Delta<0$), as we also
have seen in the calculation of the spectrum. This means that the energy provided to
the qubit is less than its excitation energy, and thus energy is taken from
the motion of the resonator to fully excite the qubit. Consequently the resonator will be cooled.
In addition we notice that the highest cooling rate is determined by a resonance
condition $\bar\Delta=\omega_{\rm m}$ as required by energy conservation. This is clearly
visible from the dotted line in the $\Delta-\Omega$-plane of figure \ref{PlotFigure1}(a). A similar resonance condition has been observed for an $LC$-oscillator
inductively coupled to a flux qubit in \cite{Hauss2008}. This result is also obvious from the analytical derivation of the spectrum
(\ref{Eq:alphas}) performed in \ref{AppC}.
Thus there is a trade-off between the optimal detuning
and the optimal drive strength. Furthermore it is clear from figure \ref{PlotFigure1}(a) that the region of optimal cooling covers only a small section of
the dotted line $\bar\Delta=\omega_{\rm m}$. This motivates us to introduce the angle $\varphi$ defined by
\begin{equation}
\sin\varphi = \Omega/\bar\Delta\label{sinvarphi}.
\end{equation}
We find that the highest cooling rate ($W>8$ kHz) appears in a region around
$\sin\varphi\approx 0.85$, which is confirmed analytically below. This means that
$\Omega$ has to be remarkably different from zero in order to get a high cooling rate,
i.e. we have to drive the system strongly. We also note that we cannot cool the resonator
below its thermal occupation for $\Delta=0$ or for $\Omega=0$. The latter is easily
understood since we need to drive the qubit in order to bring it into the excited state
from where it can decay. For $\Delta=0$ the qubit is driven at resonance, and thus will
not take energy from the resonator motion. This can also be understood from the
 the dotted spectrum in figure \ref{Fig:3Omega}, where the positive and
negative frequency parts are equally high; the spectrum (\ref{Eq:alphas}) gives ${\rm
Re}\left\{S(\pm\omega_{\rm m})\right\}=1/(3\Gamma)$ for $\Delta = 0$.

\begin{figure}
\begin{centering}
\includegraphics[width=0.8\textwidth]{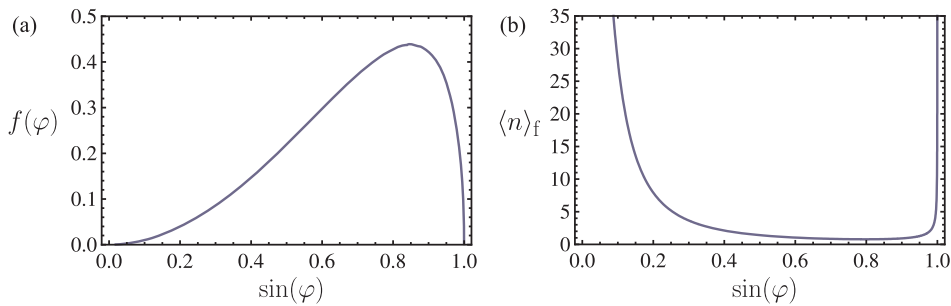}
\caption{ (a) The dimensionless function $f(\varphi)$ plotted as function of
$\sin\varphi=\Omega/\omega_{\rm m}$ with $\Omega^2 + \Delta^2 = \omega_{\rm m}^2$. The
maximum is obtained for $\sin\varphi=0.85\omega_{\rm m}$. (b) The final occupation number
as a function of the drive strength $\sin\varphi = \Omega/\omega_{\rm m}$ for a specific
set of parameters: $\omega_{\rm m}=10$ MHz: $g=0.1$ MHz, $\gamma_{\rm m}= 17$ Hz, $\Gamma
= 1$ MHz, $N_{\rm m} = 195$. The minimum is in this specific case achieved for
$\sin\varphi = 0.80$, with the minimum occupation number 0.8. \label{Fig:Womega} }
\end{centering}
\end{figure}

Now we study the cooling rate in more detail analytically, in the limit of weak qubit
decay and optimal detuning ($\Gamma\ll\bar{\Delta}=\omega_{\rm m}$). Inserting the
spectral amplitudes (\ref{Eq:alphas}) in the cooling equation (\ref{CoolingEq}), we
derive the total cooling rate $W$,
\begin{equation}
W = A_- - A_+  = 2\gamma_{\rm m} \left[ 1 + \beta  f(\varphi) \right], \label{Eq:WForOpt}
\end{equation}
with the dimensionless parameter $\beta$,
\begin{equation}
\beta = {g^2 \over \gamma_{\rm m} \Gamma}, \label{Eq:DefOfBeta}
\end{equation}
and the function $f(\varphi) = \left[4 \sin^2\varphi \sqrt{1 -
\sin^2\varphi}\right]/\left[ 4 - \sin^4 \varphi\right]$ which is plotted as a function of
$\sin\varphi$ in figure \ref{Fig:Womega}(a).
The function $f(\varphi)$ describes the behaviour along the dotted line in figure
\ref{PlotFigure1}(a). The analytical expression for the cooling rate (\ref{Eq:WForOpt})
tells us that fast cooling is obtained for a large value of the parameter $\beta$ and for
a drive strength $\Omega$ which maximizes the function $f(\varphi)$: $f_{\rm max} =
f(\Omega_{\rm 0,W})$ with $\Omega_{\rm 0,W} = 0.85\ \omega_{\rm m}$. Following the dotted
line in figure \ref{PlotFigure1}(a) and the plot of the function $f(\varphi)$ in figure
\ref{Fig:Womega}(a), we see that the cooling rate increases very slowly for weak drive,
\begin{equation}
W \simeq 2\gamma_{\rm m} \left[1 + \beta \left(\Omega \over \Delta\right)^2
\right],\qquad \Omega \ll |\Delta|
\end{equation}
up to the maximum point $\Omega_{\rm 0,W}$, from which it quickly drops to its minimum
value $W = 2 \gamma_{\rm m}$ for very strong drive,
\begin{equation}
W \simeq 2\gamma_{\rm m} \left[ 1 + \beta \left(4 |\Delta| \over 3 \Omega \right)
\right], \qquad \Omega \rightarrow \omega_{\rm m} \quad (|\Delta| \rightarrow 0).
\end{equation}

Figure \ref{PlotFigure1}(b) shows the final occupation number $\langle n\rangle_{\rm f}$
as a function of the detuning $\Delta$ and the drive strength $\Omega$ for an initial
thermal occupation of $N_{\rm m}=195$. The minimum final occupation number $\langle
n\rangle_{\rm f}<0.85$ for this specific choice of parameters is also obtained for the
detuning $\bar\Delta = \omega_{\rm m}$ which follows from energy conservation represented
by the dotted line. The optimal drive strength is in the region where $\sin\varphi\approx
0.8$. Although the final occupation number is not exactly proportional to the inverse of
the cooling rate, for the parameters we have chosen here the minimum of $\langle
n\rangle_{\rm f}$ almost coincides with the maximum of the cooling rate, which is
promising for efficient cooling procedures.

Now we study the final occupation number analytically in the limit
$\Gamma\ll\bar{\Delta}=\omega_{\rm m}$. By combining the spectrum (\ref{Eq:alphas}) with
the cooling equation (\ref{CoolingEq}) we derive the final occupation number $\langle
n\rangle_{\rm f}$,
\begin{equation}
\langle n\rangle_{\rm f} = {A_+ \over W} = {N_{\rm m} w_+(\varphi) + \beta \alpha_- \over
w_+(\varphi) + \beta (\alpha_+ - \alpha_- )}, \qquad w_+(\varphi) = {2 + \sin^2\varphi
\over 2}. \label{Eq:NF}
\end{equation}
From the expression (\ref{Eq:NF}) it is clear that for zero driving ($\sin\varphi = 0$),
the final occupation is equal to the initial (thermal) population, $\langle n\rangle_{\rm
f} = N_{\rm m}$, and thus there is neither cooling nor heating. For zero detuning
($\sin\varphi = 1$) the driven process heats up the system and finds a steady state at
$\langle n\rangle_{\rm f} = N_{\rm m} + \beta/3$. Figure \ref{Fig:Womega}(b) shows
$\langle n \rangle_{\rm f}$ as a function of $\sin \varphi = \Omega/\bar\Delta$ for a
specific set of parameters and thus gives a feeling of the qualitative behaviour.
Following figure \ref{Fig:Womega}(b) or the dotted line in figure \ref{PlotFigure1}(b),
we see that the final occupation number decreases slowly for weak drive,
\begin{equation}
\langle n\rangle_{\rm f} \simeq {N_{\rm m} \over 1 + \beta \left( \Omega \over \Delta
\right)^2}, \qquad \Omega \ll |\Delta|
\end{equation}
and finds a minimum close to the maximum for the cooling rate (here at $\Omega_{\rm 0,Nf}
= 0.80 \ \omega_{\rm m}$) after which it quickly increases again,
\begin{equation}
\langle n \rangle_{\rm f} \simeq {N_{\rm m} + \beta/3 \over 1 + \beta \left( 4 |\Delta|
\over 3 \Omega \right)},\qquad |\Delta|\to 0.
\end{equation}
In contrast to the optimum drive strength $\Omega_{\rm 0,W}$ for the
cooling rate $W$, the drive strength $\Omega_{\rm 0,Nf}$ which minimizes the final occupation
number depends on the parameters $N_{\rm m}$ and $\beta$.

\begin{figure}[h]
     \begin{center}
         \includegraphics[scale=0.6]{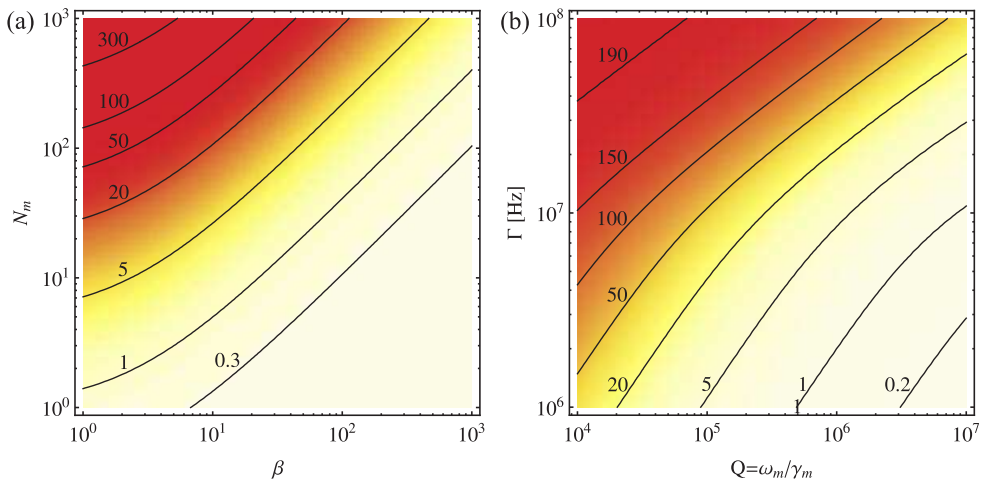}
     \end{center}
\caption{ Contour plot of the final occupation number (a) as a function of the universal
parameter $\beta=g^2/(\gamma_{\rm m}\Gamma)$ and the initial occupation number of the
nanorod $N_{\rm m}$ and (b) as a function of the quality factor $Q$ of the nanorod and
the decay rate $\Gamma$ of the qubit for the following choice of parameters: $N_{\rm
m}=195$, $g=0.1$ MHz. In both plots each point has been optimized for the detuning
$\Delta$ and the drive strength $\Omega$ . \label{PlotFigure2}}
\end{figure}

Assuming optimized drive strength $\Omega_{\rm 0,Nf}$, the thermal population $N_{\rm m}$
and $\beta$ are the {\it only} two parameters determining the final occupation number.
These parameters represent two competing processes; the intrinsic dissipation of the
resonator represented by $N_{\rm m}$ and $\gamma_{\rm m}$, and the driven cooling of
strength $g^2/\Gamma$. For the latter process to be dominant, it is required that $\beta$
is significantly larger than $N_{\rm m}$, while respecting the basic assumptions
$g/\Gamma \ll 1$ and $\gamma_{\rm m}N_{\rm m}\ll\Gamma$ of the derivation.
In figure \ref{PlotFigure2} we study the behaviour of the final occupation number beyond
the limit of resolved sidebands. Figure \ref{PlotFigure2}(a) shows how the minimized
final occupation number changes with the thermal resonator occupation $N_{\rm m}$ and the
universal parameter $\beta=g^2/(\gamma_{\rm m}\Gamma)$. We clearly see that the larger
temperature of the resonator (i.e. larger $N_{\rm m}$), the larger $\beta$ is required to
reach a low final occupation number. A larger $\beta$ is obtained {\it either} by
stronger resonator-qubit coupling $g$ {\it or} by weaker decay rates $\Gamma$ and
$\gamma_{\rm m}$. For $N_{\rm m} = 195$, corresponding to $\omega_{\rm m} =$ 10 MHz at
$T=$15 mK, ground state cooling is possible provided that $\beta > 500$. 
In figure \ref{PlotFigure2}(b) we study the final occupation number as a function of the
quality factor $Q=\omega_{\rm m}/\gamma_{\rm m}$ of the resonator and the decay rate
$\Gamma$ of the qubit, for a specific thermal occupation number $N_{\rm m}=195$ and fixed
coupling $g=0.1$ MHz. A lower occupation number is achieved by increasing the quality
factor of the resonator, or by decreasing the decay rate of the qubit. For $\Gamma= 1$
MHz ground state cooling can be obtained for $Q>5\cdot 10^5$, which corresponds to
$\gamma_m < 20$ Hz for $\omega_{\rm m}=10$ MHz. This corresponds to the resolved sideband
limit and the parameters we used so far.

Remember that the results are presented in a frame where the resonator motion is shifted
by a coherent shift $\alpha$ (\ref{Eq:ShiftAlpha}). In the original frame we would have
to add the coherent part $\sim |\alpha|^2$ to $\langle n\rangle_{\rm f}$. This shift can
be removed by adiabatically switching off the coupling magnetic field $B_0$ or the
circulating currents $I_{\rm q}$.

The results presented in this discussion are valid in the limit of negligible pure dephasing, $\Gamma_{\rm d}/\Gamma \rightarrow 0$.
For finite $\Gamma_{\rm d}$ one has to consider the more general analytic expressions in \ref{AppC} for the cooling rate $W$ (\ref{Eq:WForOptApp}) and the final occupation number $\langle n \rangle_{\rm f}$ (\ref{Eq:NFApp}), which are valid for $\Gamma,\Gamma_{\rm d} \ll \bar\Delta$. From these expression we draw the conclusion that
our results are still valid for weak dephasing rate $\Gamma_{\rm d}\ll\Gamma$, whereas a large $\Gamma_{\rm d}$ would change the conditions for ground-state cooling considerably.

As a last remark, we point out the similarities and differences of the
 cooling process discussed in this paper compared to the laser cooling of a trapped ion
\cite{Stenholm1986}. The dynamics of the two-level system in this setup (the CPB qubit)
is formally equivalent (in the limit $\Gamma_d/\Gamma \rightarrow 0$) to the damped and
driven two level system modeling the ion, with its dynamics completely described by the
Bloch equation (\ref{Eq:Bloch}). The position of the peaks in the spectrum is fully
determined by the eigenvalues of the matrix $A$ (\ref{Eq:AMatrix}). This explains why we
obtain optimal cooling in the same regime, where cooling for a trapped ion is optimal,
namely in the resolved sideband regime. The main difference between the two systems is
the cooling force, which in this setup gives an interaction of the form $\sim\sigma_z(a +
a^\dag)$, in contrast to the Jaynes-Cummings term $\sim (\sigma_+ a + \sigma_- a^\dag)$
which appears for ion cooling. This difference manifests itself in the dependence on
$\Delta$ and $\Omega$ of the amplitude and the effective width of the spectral peaks, and
consequently affects the optimal drive strength. As already mentioned earlier, there are
contributions to the cooling rate both from the intrinsic dissipation $\gamma_{\rm m}$ of
the resonator and through the coupling to the two-level system, in contrast to ion
cooling where only the latter is present since the motion of the ion is perfectly
isolated. Finally we should mention that whereas the laser both drives the internal
degrees of freedom of the ion and its coupling to the motion, in this setup the qubit
coupling to the motion is independent from the microwave signal which drives the qubit.

\section{Conclusion\label{Conclusion}}

In this paper we have derived the conditions for ground-state cooling of a nanoresonator
beam which is integrated in a loop-shaped Cooper-pair box (CPB) circuit. The CPB qubit
acts as leaky two-level system through which the energy of the resonator motion
dissipates. The cooling force in question, the qubit-state dependent Lorentz force, acts
on the beam due to circulating currents in the circuit in combination with an applied
magnetic field. We have derived a general expression for the force spectrum, which
determines the cooling rate and the final occupation number, and evaluated the result in
the resolved sideband regime where the qubit decay $\Gamma$ is small compared to the
resonator frequency $\omega_{\rm m}$.

In the resolved sideband regime, the spectrum exhibits narrow peaks and cooling is
optimized for the detuning $\sqrt{\Delta^2 + \Omega^2} = \omega_{\rm m}$. Assuming this
condition, we further investigated the cooling rate and the final occupation number as a
function of the drive strength $\Omega$, and found that optimal cooling requires a
strongly driven qubit; the cooling rate is maximal for $\Omega = 0.85 \ \omega_{\rm m}$
independently of the other parameters. The final occupation number is minimized in the
region $\Omega \simeq 0.8 \ \omega_{\rm m}$ for our parameter example. When optimizing
the final occupation number with respect to the drive strength, we found that basically
two parameters determine the possibility to cool the resonator mode to the ground state,
namely the initial thermal occupation number $N_{\rm m}$ of the resonator, and the
parameter $\beta = g^2/(\Gamma \gamma_{\rm m})$. Thus given an initial occupation number
of the mechanical mode, the possibility to reach the ground state is determined by the
relation between the coupling strength $g$ and the decay rates $\Gamma$ and $\gamma_{\rm
m}$, requiring strong enough coupling in comparison to the decay rates. Generally cooling
requires $\beta > N_{\rm m}$, and in particular we could show that for $N_{\rm m}=195$,
corresponding to $\omega_{\rm m}= 10$ MHz at $T=15$ mK, ground state cooling requires
$\beta > 500$.

Let us also say a few words about the limiting cases. For $\Omega = 0$, the qubit is not
driven and provides neither cooling nor heating. The cooling rate is given by the
dissipation rate of the resonator, $2 \gamma_{\rm m}$, and thus the mechanical mode stays
at thermal occupation $\langle n \rangle_{\rm f} = N_{\rm m}$. For zero detuning, $\Delta
= 0$, the qubit provides both cooling and heating of the resonator motion with equal
amplitude, and the resonator is heated up to the final occupation number $N_{\rm m} +
\beta/3$. The total cooling rate is still given by  $2 \gamma_{\rm m}$.

From the form of the parameter $\beta$ it is clear that for a specific coupling strength
$g$, cooling is improved by a larger $Q$ of the resonator, which basically reflects the
fact that the resonator environment needs longer time to heat up the mechanical mode to
its thermal state after it has been cooled.

Cooling is also improved by a weaker qubit decay $\Gamma$, which is not very intuitive
since the cooling procedure relies on the dissipation of the excited qubit state.
However, from the point of view of the resonator, the decay of the qubit state shows up
as an effective decay rate $g^2/\Gamma$ in for example the heating and cooling rates
$A_\pm$, and therefore the cooling efficiency decreases with larger $\Gamma$.

\section*{Acknowledgement}

We acknowledge interesting and useful discussions with V. Shumeiko, A. Shnirman, P. Zoller and M. M\"uller, and help with the figure files from A. Micheli.
This work was supported by the Austrian FWF through SFB F15, and the EU networks CONQUEST
and EuroSQIP.

\appendix


\section{Derivation of the Hamiltonian for the Cooper pair box \label{AppA}}

In the following we derive the qubit Hamiltonian (\ref{HTLS}) for the CPB circuit shown in
figure \ref{SetupFigure}(b). Let us start from the circuit Lagrangian
$L(\dot\phi_i,\phi_i)$ \cite{Devoret1997}, which is presented in terms of the phases
$\phi_1,\phi_2,\phi$ across the respective Josephson junctions (JJs) of the circuit,
using the voltage-phase relation $V=(\hbar/(2e))\dot\phi$,
\begin{equation}
L = L_{\rm SCB} + L_{\rm JJ},
\end{equation}
where the first term describes the driven Cooper pair box,
\begin{equation}
\fl L_{\rm SCB} =
\left(\frac{\hbar}{2e}\right)^2\left[\frac{C}{2}(\dot{\phi}_1^2+\dot{\phi}_2^2)
+\frac{C_{\rm g}}{2}\left(\frac{2e}{\hbar}V_{\rm g}-\dot{\phi}_1\right)^2\right]
+E_{\rm J}(\cos\phi_1+\cos\phi_2)
\end{equation}
and the second term describes the large current-biased Josephson junction (JJ),
\begin{equation}
L_{\rm JJ} = \left(\frac{\hbar}{2e}\right)^2 {C_{\rm b} \over 2} \dot\phi^2 +
E_{\rm J,b} \cos\phi+\frac{\hbar}{2e}I_{\rm b}\phi .
\end{equation}
In the following we assume the phase regime $E_{\rm J,b} \gg E_{\rm C,b} = (2e)^2/(2C_{\rm b})$ for the
large JJ, in which case the dynamics of the phase $\phi$ is similar to that of a particle in a
tilted washboard potential, with the tilt determined by the applied bias current
$I_{\rm b}$. Provided that the tilt is not too large, $I_{\rm b}< I_{\rm c}$ where $I_{\rm c} = (2e/\hbar)E_{\rm J,b}$
is the critical current of the JJ, the potential contains well-defined wells in
which the phase is trapped. Further we assume that $C_{\rm b} \rightarrow \infty$,
corresponding to a particle with infinite mass, in which case the phase stays at the
minimum point $\phi_0$ determined by the bias current via the relation
$\sin{\phi_0}=I_{\rm b}/I_{\rm c}$, and the large JJ can be treated as a classical device without
dynamics. Here we also assumed that the currents through the large JJ due to the CPB
dynamics are small, which is justified if $E_{\rm J}/E_{\rm J,b} \ll 1$. The
contribution from the large JJ to the Lagrangian is thus only a constant term $L_{\rm JJ} (\phi =\phi_0)$. On
the other hand, assuming negligible self-inductance in the loop, the phase $\phi_0$
determines the phase across both CPB junctions through the flux quantization relation \cite{Tinkham1996},
\begin{equation}
\phi_1+\phi_2+\phi_0=0.
\end{equation}
A similar effect is achieved by applying a magnetic flux
$\Phi_{\rm e}$ as shown in figure \ref{SetupFigure}(b). In the following it is
understood that $\phi_0$ either is determined by a bias current or by an applied magnetic
flux with $\phi_0 = (2e/\hbar)\Phi_{\rm e}$. We are left with only one degree of
freedom in the system,
\begin{equation}
\phi_{\rm q} = \frac{\phi_2-\phi_1}{2},
\end{equation}
and the circuit Lagrangian gets the form,
\begin{equation}
L_{\rm SCB} = \left(\frac{\hbar}{2e}\right)^2\left[C\dot{\phi}_{\rm q}^2 +\frac{C_{\rm
g}}{2}\left(\frac{2e}{\hbar}V_{\rm g}+\dot{\phi}_{\rm q}\right)^2\right] +2E_{\rm
J}\cos{\left(\frac{\phi_0}{2}\right)}\cos{\phi_{\rm q}} .
\end{equation}
In order to derive the circuit Hamiltonian, we calculate the charge number on the island
$n_{\rm q}$, which is the conjugate variable to the phase $\phi_{\rm q}$,
\begin{equation}
n_{\rm q} = \frac{1}{\hbar}\frac{\partial L}{\partial\dot\phi_{\rm q}},
\end{equation}
and obtain the circuit Hamiltonian $H (n_{\rm q},\phi_{\rm q}) = n_{\rm q} \dot\phi_{\rm q} - L$ \cite{Goldstein},
\begin{equation}
H = E_{\rm C}\left(n_{\rm q}-n_{\rm g} \right)^2 - 2E_{\rm J}\cos{\left(\frac{\phi_0}{2}\right)}\cos{\phi_{\rm q}}
\label{HnachL}
\end{equation}
where the charging energy $E_{\rm C}=(2e)^2/(2C_{\Sigma})$ contains the total island
capacitance $C_\Sigma=2C+C_{\rm g}$, and $n_{\rm g} = C_{\rm g} V_{\rm g}/(2e)$ is the
dimensionless charge induced on the island by the gate voltage. Note here that the
two-junction circuit effectively behaves as a single junction CPB with tunable Josephson
energy $2E_{\rm J}\cos(\phi_0/2)$. Now we can also derive the current flowing in the
circuit. The loop current is determined by the time derivative of the charge $q$ through
both JJs, which is conjugate to the phase $-(\phi_1+\phi_2)=\phi_0$ across the CPB. Using
Hamilton's equations of motion \cite{Goldstein}, we get,
\begin{equation}
I = \dot{q} = {2e \over \hbar} \frac{\partial H}{\partial \phi_0} = 2 I_{\rm q}
\cos\phi_{\rm q} \label{Eq:loopI}
\end{equation}
with the current $I_{\rm q}$ given by (\ref{Eq:Iq}).
Further we quantize the circuit Hamiltonian (\ref{HnachL}) \cite{Devoret1997},
\begin{equation}
H = E_{\rm C}\left(\hat{n}_{\rm q}-n_{\rm g} \right)^2 - 2E_{\rm J}\cos{\left(\frac{\phi_0}{2}\right)}\cos{\hat{\phi}_{\rm q}}
\label{Eq:Hquant}
\end{equation}
by imposing the canonical commutation relation $[\hat{\phi}_{\rm q},\hat{n}_{\rm q} ]=i$.
In the absence of tunneling ($E_{\rm J} \rightarrow 0$), the charge states $|2en\rangle$
representing integer number of Cooper pairs on the island form an eigenbasis for the
Hamiltonian (\ref{Eq:Hquant}). For dominating charging energy, $E_{\rm C} \gg E_{\rm J}$,
the two lowest energy eigenstates are superpositions of the two charge states
$|2en\rangle$ and $|2e(n+1)\rangle$ \cite{Shnirman1997}. The superposition states are
controlled by tuning the dc-part of the gate voltage $V_{\rm g,0}$. In particular, we can
bias the island at the charge degeneracy point $n_{\rm g}=1/2$ where $|2en\rangle$ and
$|2e(n+1)\rangle$ are degenerate and the energy levels are split up by the weak
tunneling. The eigenstates are given by the symmetric and antisymmetric superposition
states $|e/g \rangle=\left(|2en \rangle\pm |2e(n+1)\rangle\right)/\sqrt{2}$. Since the
spectrum is highly asymmetric, it is allowed to consider only these two levels and treat
the system as a qubit \cite{Shnirman1997}, provided that the temperature is sufficiently
low. At the charge degeneracy point the truncated Hamiltonian for the qubit reads,
\begin{equation}
H = {E \over 2}\sigma_z + \hbar\Omega(t) \sigma_x,\label{qBnachL}
\end{equation}
written in terms of usual Pauli matrices in the basis $\{|e\rangle, |g\rangle \}$. Here
we assume the energy splitting $E = 2 E_{\rm J} \cos(\phi_0/2)$ to be tuned
adiabatically, whereas $\Omega(t) = \left(E_{\rm C}/\hbar\right) (n_{\rm g} (t) - 1/2)$
is driven by the gate voltage; $n_{\rm g}(t) = 1/2 + \delta n_{\rm g} \cos\omega_{\rm
dr}t$. Moving to a frame rotating with the microwave frequency $\omega_{\rm dr}$, we end
up with the qubit Hamiltonian (\ref{HTLS}).

During quantization and truncation, the expression for the loop current (\ref{Eq:loopI})
transforms into a current operator which is diagonal in the qubit eigenbasis
$\{|e\rangle,|g\rangle\}$ \cite{Vion2002},
\begin{equation}
\hat{I} = I_{\rm q} \sigma_z .
\label{Eq:Ioperator}
\end{equation}
This holds also in the regime $E_{\rm C} \sim E_{\rm J}$ \cite{Wallquist2005,WallquistThesis}.

%
%

So far we have assumed the ideal case of a symmetric Cooper-pair box circuit, i.e. one
where the two JJs are equally large. In reality the deviation $\delta E_{\rm J} = (E_{\rm
J,1}- E_{\rm J,2})$ is usually a few percent, $\delta E_{\rm J} \ll E_{\rm J}$. We expand
the qubit Hamiltonian $H$ (\ref{qBnachL}) to first order in $\delta E_{\rm J}/E_{\rm J}$,
\begin{equation}
H \rightarrow {E \over 2} \sigma_z + \hbar\Omega(t) \sigma_x + {\delta E \over 2}
\sigma_y, \label{Eq:Hqubit}
\end{equation}
with $\delta E = \delta E_J \sin (\phi_0/2)$. Further we investigate the effect of the
asymmetry on the qubit-resonator interaction (\ref{Eq:Hint}). In addition to the excited
loop current $\hat{I}$ (\ref{Eq:Ioperator}), the asymmetry creates an additional
qubit-resonator interaction term via the current $\hat{I}_{\rm a}$
\begin{equation}
\hat{I}_{\rm a} = I_{\rm a} \sigma_y \quad {\rm with} \qquad
I_{\rm a} = {\delta E_J \over 4} \cos {\phi_0 \over 2} , \label{Eq:asymI}
\end{equation}
which is non-zero even for zero bias current ($\phi_0=0$).
However, in the rotating frame of equations (\ref{HTLS}) and (\ref{Eq:Hint}), the asymmetry terms are rotating fast,
\begin{equation}
\sigma_y \rightarrow i \left[ e^{-i\omega_{dr}t} \sigma_- - e^{i\omega_{dr}t} \sigma_+
\right],
\label{Eq:sigmay}
\end{equation}
and thus within the rotating wave approximation, a small
asymmetry $\delta E_J/E_J \ll 1$ does not affect the qubit dynamics or the qubit-resonator interaction.


\section{Master Equation\label{AppB}}

In this section we derive an effective master equation for the motion of the resonator
starting from the master equation for the full system given in equations
(\ref{MasterEq})-(\ref{MasterEqL2}). First we perform an adiabatic elimination of the
qubit degrees of freedom assuming $\Gamma\gg g$ such that the qubit goes to a steady
state on a much shorter time scale than the time scale for the interaction of the
resonator with the qubit. Thus we can assume that the density operator $\rho$ never
deviates much from the factorized form $\rho_{\rm ss} \otimes \tr_{\rm q}\{\rho \}$ where
the qubit is in its steady state $\rho_{\rm ss}$, and project the master equation onto
this form. To remove the action of the coupling term on the projected state, we begin by
defining a displacement operator $D$ which performs a coherent shift of the oscillator
displacement,
\begin{eqnarray}
DaD^\dag&=&a+\alpha.\label{Eq:Shifta}
\end{eqnarray}
$D$ is a unitary operator which acts only in the resonator space. We define the density
operator $\bar{\rho}$ in the shifted frame,
\begin{eqnarray}
\bar{\rho}&=&D\rho D^\dag,\label{Eq:Shiftrho}
\end{eqnarray}
and similarly applies the displacement operator to the master equation,
\begin{eqnarray}
\dot{\bar{\rho}}&=&\mathcal{L}_0^{\rm q}\bar\rho + D(\mathcal{L}_0^{\rm m}\rho+\mathcal{L}_1\rho+\mathcal{L}_2\rho)D^\dag
\end{eqnarray}
with
\begin{eqnarray}
 D\mathcal{L}_0^{\rm m}\rho D^\dag&=& \mathcal{L}_0^{\rm m}\bar{\rho}-i\omega_{\rm m}[\alpha a^\dag+\alpha^*
a,\bar{\rho}],\nonumber\\
D\mathcal{L}_1\rho
D^\dag&=&\mathcal{L}_1\bar{\rho}-ig(\alpha+\alpha^*)[\sigma_z,\bar{\rho}],\\
D\mathcal{L}_2\rho D^\dag&=&\mathcal{L}_2\bar{\rho}+\gamma_{\rm m}[\alpha^*a-\alpha
a^\dag,\bar{\rho}].
\end{eqnarray}
The shift acquired in $D\mathcal{L}_1D^\dag$ acts only on the qubit, and we include
it in the qubit dynamics by defining a new Liouvillian $\tilde{\mathcal{L}}_0^{\rm q}$,
\begin{eqnarray}
\tilde{\mathcal{L}}_0^{\rm q}\bar{\rho}&=&\mathcal{L}_0^{\rm q}\bar{\rho}-ig(\alpha+\alpha^*)[\sigma_z,\bar{\rho}].
\end{eqnarray}
Then we can write the master equation in the shifted frame as,
\begin{eqnarray}
\dot{\bar{\rho}}&=&(\mathcal{L}_0^{\rm m}+\tilde{\mathcal{L}}_0^{\rm q}+\mathcal{L}_1+\mathcal{L}_2)\bar{\rho}
+[(\gamma_{\rm m}-i\omega_{\rm m})\alpha^*a-(\gamma_{\rm m}+i\omega_{\rm m})\alpha
a^\dag,\rho].\label{MEprojected}
\end{eqnarray}
The next step is to project the master equation onto the factorized form,
\begin{eqnarray}
P\bar{\rho}=\rho_{\rm ss}\otimes \bar{\rho}_{\rm m},\qquad \bar{\rho}_{\rm m}={\rm
tr}_{\rm q}\left\{\bar{\rho}\right\}\label{PProjectionDef}
\end{eqnarray}
using the projection operator $P$ defined by $P\tilde{\mathcal{L}}_0^{\rm q}\bar{\rho}=0$
and the projection on the orthogonal subspace $Q=1-P$.
The shift $\alpha$ is determined by requiring that the projection on $P$ of the last
term in (\ref{MEprojected}) cancels the projected coupling term
$P\mathcal{L}_1P\bar{\rho}=-ig\rho_{\rm ss}\otimes \langle
\sigma_z\rangle[a+a^\dag,\bar{\rho}_{\rm m}]$,
\begin{eqnarray}
\alpha&=&-\frac{g\langle \sigma_z\rangle}{\omega_{\rm m}-i\gamma_{\rm m}}.
\label{Eq:ShiftAlpha}
 \end{eqnarray}
The master equation in the $P$-space now reads,
\begin{eqnarray}
P\dot{\bar{\rho}}&=&P\mathcal{L}_0^{\rm m} P\bar{\rho}+P\mathcal{L}_1Q\bar{\rho}+P\mathcal{L}_2P\bar{\rho}\label{PProjection1},
\end{eqnarray}
where we made use of $[P,\mathcal{L}_0^{\rm m}]=0$ and $[P,\mathcal{L}_2]=0$.
Remember that our goal is to derive a master equation for $\bar{\rho}_{\rm m}$ which is
correct up to the second order in the small parameters $\epsilon_{1(2)}$, as we outlined in section \ref{CoolingSection}. The procedure is to derive
a closed equation for $P\bar{\rho}$. We first have to replace
$Q\bar{\rho}$ in (\ref{PProjection1}) by an expression which only depends on
$P\bar{\rho}$ and is correct up to first order in the small parameters. This is
sufficient as $Q\bar{\rho}$ in (\ref{PProjection1}) is multiplied by $\mathcal{L}_1$,
which is already of order $O(\epsilon_1)$.
Using $Q\mathcal{L}_0^{\rm m}P\bar{\rho}=0$, $Q\mathcal{L}_0^{\rm q}P\bar{\rho}=0$ and
$Q\mathcal{L}_2P\bar{\rho}=0$
we write the master equation in the $Q$-space
\begin{eqnarray}
\fl
Q\dot{\bar{\rho}}&=&Q \left( \mathcal{L}_0^{\rm m} + \tilde{\mathcal{L}}_0^{\rm q} + \mathcal{L}_2 \right) Q\bar{\rho}+Q\mathcal{L}_1P\bar{\rho}
-ig  Q[\left(\sigma_z - \langle\sigma_z\rangle \right)(a+a^\dag),Q\bar{\rho}].
\label{QProjection1}
\end{eqnarray}
As the term $Q\bar{\rho}$ is of order $O(\epsilon_1)$, the last term in
(\ref{QProjection1}) is of order $O(\epsilon_1^2)$ and the term
$Q\mathcal{L}_2Q\bar{\rho}$ is of order $O(\epsilon_1\epsilon_2)$. Thus we can neglect
these terms and solve the equation formally,
\begin{eqnarray}
Q\bar{\rho}(t)&=&e^{Q(\mathcal{L}_0^{\rm m}+\tilde{\mathcal{L}}_0^{\rm q})(t-t_1)}Q\bar{\rho}(t_1)+\int\limits_{t_1}^t
d\tau
e^{Q(\mathcal{L}_0^{\rm m}+\tilde{\mathcal{L}}_0^{\rm q})(t-\tau)}Q\mathcal{L}_1P\bar{\rho}(\tau).
\label{QProjSol1}
\end{eqnarray}
This equation is exact up to $O(\epsilon_{1(2)})$. Next we solve the equation for
$P\bar{\rho}$ (\ref{PProjection1}) in order to insert it into the term
$Q\mathcal{L}_1P\bar{\rho}$ in equation (\ref{QProjSol1}). To be consistent we have to
keep terms only up to order $O(\epsilon_{1(2)}^0)$,
\begin{eqnarray}
P\bar{\rho}(t)&=&e^{P\mathcal{L}_0^{\rm m}(t-t_0)}P\bar{\rho}(t_0)\label{PProjSol1}.
\end{eqnarray}
Being interested in timescales larger than $1/\Gamma$ for which
$\tilde{\mathcal{L}}_0^{\rm q}$ decays, the first term in (\ref{QProjSol1})
will not contribute to the master equation. The integrand in the second term of (\ref{QProjSol1}) decays on
a time scale $1/\Gamma$, allowing a Markov approximation where we extend the
lower limit of integration to $t_1\to-\infty$ and replace $\bar\rho(t_0)$ by $\bar\rho(t)$. Inserting the solution for $P\bar{\rho}(t)$ (\ref{PProjSol1}) into
the solution for $Q\bar{\rho}(t)$ (\ref{QProjSol1}) we obtain,
\begin{eqnarray}
Q\bar{\rho}(t)&=&\int\limits_{0}^{\infty} d\tau
e^{Q(\mathcal{L}_0^{\rm m}+\tilde{\mathcal{L}}_0^{\rm q})\tau}Q\mathcal{L}_1
e^{-P\mathcal{L}_0^{\rm m}\tau}P\bar{\rho}(t). \label{QProjSol2}
\end{eqnarray}
Further inserting this solution for $Q\bar{\rho}(t)$ (\ref{QProjSol2}) into (\ref{PProjection1}), we finally obtain a closed equation for
$P\bar{\rho}$. Using the definition for $P$ (\ref{PProjectionDef}) and tracing over the qubit degree of freedom, the master equation for the resonator motion
$\bar{\rho}_{\rm m}$ is expressed as,
\begin{eqnarray}
\dot{\bar{\rho}}_{\rm m}&=& (\mathcal{L}_0^{\rm m}+\mathcal{L}_2)\bar{\rho}_{\rm m} +{\rm
tr}_{\rm q}\left\{
 \int\limits_{0}^{\infty}
d\tau \mathcal{L}_1Q e^{(\mathcal{L}_0^{\rm m}+\tilde{\mathcal{L}}_0^{\rm q})\tau}\mathcal{L}_1
\rho_{\rm ss}\otimes e^{-\mathcal{L}_0^{\rm m}\tau}\bar{\rho}_{\rm m}\right\}.
\label{eq:ME}
\end{eqnarray}
There is still some work to do to evaluate the second term of the master equation for the resonator (\ref{eq:ME}).
We define the operator,
\begin{equation}
\delta\sigma_z = \sigma_z-\langle\sigma_z\rangle_{\rm ss},
\end{equation}
describing fluctuations of $\sigma_z$ about its mean steady state value
$\langle\sigma_z\rangle_{\rm ss}$. Further we use the following relations: ${\rm tr}_{\rm
q}\left\{\delta\sigma_z e^{\tilde{\mathcal{L}}_0^{\rm q}\tau}\delta\sigma_z\rho_{\rm
ss}\right\} =  \langle \delta\sigma_z(\tau)\delta\sigma_z\rangle_{\rm ss}$, ${\rm
tr}_{\rm q}\left\{\delta\sigma_z e^{\tilde{\mathcal{L}}_0^{\rm q}\tau}\rho_{\rm
ss}\delta\sigma_z\right\}= \langle \delta\sigma_z\delta\sigma_z(\tau)\rangle_{\rm ss}$
and $e^{\mathcal{L}_0^{\rm m}\tau}a = a(-\tau) $.
After some algebra we obtain a master equation for $\bar{\rho}_{\rm m}$ where the qubit degree of freedom is completely traced out,
\begin{eqnarray}
\dot{\bar{\rho}}_{\rm m}&=& (\mathcal{L}_0^{\rm m}+\mathcal{L}_2)\bar{\rho}_{\rm m} -g^2
\int\limits_{0}^{\infty} d\tau \langle \delta\sigma_z(\tau)\delta\sigma_z\rangle_{\rm ss}
[a+a^\dag,[a(-\tau)+a^\dag(-\tau),\bar{\rho}_{\rm m}]]\nonumber\\
&-& g^2 \int\limits_{0}^{\infty} d\tau \langle
[\delta\sigma_z(\tau),\delta\sigma_z]\rangle_{\rm ss} [a+a^\dag,\bar{\rho}_{\rm
m}(a(-\tau)+a^\dag(-\tau))].
\end{eqnarray}
In the following we show how to bring this equation into the form (\ref{EQ:reducME}). We
define a spectral function $S(\omega)$ (\ref{ForceSpectrum})
as the Laplace transform of the correlation function $\langle
\delta\sigma_z(\tau)\delta\sigma_z\rangle_{\rm ss}$ of the qubit operators at the steady
state.
The steady state of the qubit should in principle be calculated with respect to
the shifted Liouvillian $\tilde{\mathcal{L}}_0^{\rm q}$, which is written using the value for $\alpha$ (\ref{Eq:ShiftAlpha}),
\begin{eqnarray}
\tilde{\mathcal{L}}_0^{\rm q}\bar{\rho} \ &=& \ \mathcal{L}_0^{\rm q}\bar{\rho}
+\frac{2ig^2\omega_{\rm m}\langle\sigma_z\rangle_{\rm ss}}{\omega_{\rm m}^2+\gamma_{\rm
m}^2}[\sigma_z,\bar{\rho}] .
\end{eqnarray}
The second term is included into the detuning $\tilde{\Delta}=\Delta+4g^2\omega_{\rm
m}/(\omega_{\rm m}^2+\gamma_{\rm m}^2)\langle\sigma_z\rangle_{\rm ss}$, but the
correction is small, $(g/\omega_{\rm m})^2\ll 1$. Therefore it is justified to
approximate the detuning by $\tilde{\Delta}\approx \Delta$ and to calculate the steady
state with respect to $\mathcal{L}_0^{\rm q}$.
The master equation for the motion of the resonator finally gets the form,
\begin{eqnarray}
\dot{\bar{\rho}}_{\rm m}&=&(\mathcal{L}_0^{\rm m}+\mathcal{L}_2^{\rm m})\bar{\rho}_{\rm m} \nonumber \\
&&-g^2 \left[ S(\omega_{\rm m})[a+a^\dag,[a,\bar{\rho}_{\rm m}]] +
S(-\omega_{\rm m})[a+a^\dag,[a^\dag,\bar{\rho}_{\rm m}]] \right]\nonumber\\
&&-g^2 ( S(\omega_{\rm m})-S^*(-\omega_{\rm m}))[a+a^\dag,\bar{\rho}_{\rm m}a] \nonumber \\
&& -g^2( S(-\omega_{\rm m})-S^*(\omega_{\rm m}))[a+a^\dag,\bar{\rho}_{\rm m}a^\dag].
\end{eqnarray}
After applying the rotating wave approximation, where we neglect the fast oscillating
terms which are an order $g^2/(\omega_{\rm m}\Gamma)$ smaller than the slow terms,
we obtain,
\begin{eqnarray}
\dot{\bar{\rho}}_{\rm m}&=&-i\omega_{\rm m}[a^\dag a,\bar{\rho}_{\rm m}]-ig^2 \left[ {\rm
Im}\left\{S(\omega_{\rm m})\right\}[a^\dag a,\bar{\rho}_{\rm m}] + {\rm
Im}\left\{S(-\omega_{\rm m})\right\}[aa^\dag,\bar{\rho}_{\rm m}] \right] \nonumber\\
&& +(g^2 {\rm Re}\{S(\omega_{\rm m})\}+\gamma_{\rm m}(N_{\rm m}+1))\ {\cal D}[a](\bar\rho_{\rm m})\nonumber\\
&&+(g^2 {\rm Re}\left\{S(-\omega_{\rm m})\right\}+\gamma_{\rm m}N_{\rm m})\ {\cal
D}[a^\dag](\bar{\rho}_{\rm m}),
\end{eqnarray}
which can also be written on the form (\ref{EQ:reducME}) by introducing the effective frequency $\tilde\omega_{\rm m}$ (\ref{Eq:omegameff}) and cooling and heating
rates $A_\pm$ (\ref{Apm}).
Note that we discuss the results in section \ref{Discussion} in the frame where the
resonator is displaced by the coherent shift $\alpha$ (\ref{Eq:ShiftAlpha}). The master
equation in the unshifted frame contains additional terms $\sim\alpha a^\dag, \alpha^* a$
which do not contribute to the cooling equation, and a term $|\alpha|^2$ which shifts the
final occupation number.


  \section{Spectrum \label{AppC}}
  In this section we first derive a general expression for the force
spectrum
  $S(\omega)$
  given in equation (\ref{ForceSpectrum}), in terms of the parameters
$\Gamma,\Gamma_{\rm d}, \Delta$ and $\Omega$ of the qubit dynamics. Further we evaluate
the real part of the force
  spectrum in
  the limit of well-resolved peaks, corresponding to the resolved
sideband limit
  for ion
  cooling \cite{Stenholm1986}. To begin, let us write the qubit dynamics
  (\ref{MasterEqL0q})
  in terms of the corresponding Bloch equation
  \begin{equation}
  \partial_t \langle \vec\sigma \rangle = A \langle \vec\sigma \rangle -
  \vec\Gamma
  \label{Eq:Bloch}
  \end{equation}
  for the Pauli operators $(\sigma_x,\sigma_y,\sigma_z)$, with
  \begin{equation}
  A = \left(
  \begin{array}{ccc}
  - \Gamma - \Gamma_d & \Delta & 0 \\
  - \Delta & - \Gamma - \Gamma_d & - \Omega \\
  0 & \Omega & - 2\Gamma
  \end{array}
  \right), \qquad \vec\Gamma = \left(
  \begin{array}{c}
  0 \\
  0 \\
  2 \Gamma
  \end{array}
  \right), \label{Eq:AMatrix}
  \end{equation}
  and steady state solution $\langle \vec\sigma \rangle_{\rm ss} =
  A^{-1}\vec\Gamma$.
  We further note that the spectrum in equation (\ref{ForceSpectrum}) is
given by
  the
  Laplace transform of the correlation function $\langle \delta \sigma_z (t)
  \delta
  \sigma_z (0)\rangle_{\rm ss}$ for the fluctuation operators $\delta
\vec\sigma =
  \vec\sigma - \langle \vec\sigma \rangle_{\rm ss}$,
  \begin{equation}
  S(\omega) = \langle \delta \sigma_z (s = -i\omega) \delta \sigma_z (0)
  \rangle_{\rm ss} ,
  \end{equation}
  which can be derived from the equation of motion for $\langle \delta
\sigma_z
  (t) \delta
  \sigma_z (0) \rangle_{\rm ss}$. In a first step we derive the equation
of motion
  for
  $\langle \delta \vec\sigma \rangle$ from the Bloch equations
(\ref{Eq:Bloch}),
  \begin{equation}
  \partial_t \langle \delta \vec\sigma \rangle = A \langle \delta \vec\sigma
  \rangle .
  \end{equation}
  Then using the quantum regression theorem \cite{WallsMilburn}, we find the
  solution for
  the spectrum,
  \begin{equation}
  S(\omega) = - (0,0,1) [i \omega {\mathbf 1} + A]^{-1} \vec{B},
\label{Eq:SR}
  \end{equation}
  with
  \begin{equation}
  \vec{B} = \langle \delta \vec\sigma \delta \sigma_z \rangle_{\rm ss} =
\left(
  \begin{array}{c}
  -i \langle \sigma_y \rangle_{\rm ss} - \langle \sigma_x \rangle_{\rm
ss} \langle
  \sigma_z \rangle_{\rm ss} \\
  i \langle \sigma_x \rangle_{\rm ss} - \langle \sigma_y \rangle_{\rm
ss} \langle
  \sigma_z \rangle_{\rm ss} \\
  1 - \langle \sigma_z \rangle_{\rm ss}^2
  \end{array}
  \right).
  \end{equation}
  In principle, we have now arrived at an expression for the force
spectrum in
  terms of the
  qubit dynamics through the matrix $A$ and the vector $\vec{B}$. Let us
proceed
  one step
  further and write the inverse matrix in the spectrum (\ref{Eq:SR}) as
$[i \omega
  {\mathbf 1} +
  A]^{-1} = {\rm Adj}[i \omega {\mathbf 1} + A]/{\rm Det}[i \omega
{\mathbf 1} +
  A]$, where
  ${\rm Adj}(x)$ denotes the adjugate matrix of $x$ and ${\rm Det}(x)$
denotes its
  determinant. The spectrum then gets the form,
  \begin{equation}
  \fl S(\omega) =  {h (\omega) \over
  (i\omega+\epsilon_0)(i\omega+\epsilon_+)(i\omega+\epsilon_-)},\qquad
h(\omega) =
  -
  (0,0,1) {\rm Adj}[i \omega {\mathbf 1} + A] \vec{B}, \label{SRApp}
  \end{equation}
  where $\epsilon_i$ are the eigenvalues of the matrix $A$. From this
last step it
  is clear
  that the spectrum generally exhibits three peaks, with position and
width given
  respectively by the imaginary and real parts of the eigenvalues of $A$.

  For our purposes it is sufficient to look at the real part of the spectrum
  $S(\omega)$, as the imaginary part only contributes a small shift to $\omega_{\rm m}$,
  but does not contribute to the cooling.
  In the following
  we focus on the case of well-resolved peaks, which intuitively assumes
weak
  qubit decay
  $\Gamma,\Gamma_{\rm d}$. With a spectrum containing only a few narrow peaks,
cooling is
  significant only
  when the peaks are situated at the trap frequency $\omega_{\rm m}$, and we
  therefore
  evaluate the spectrum only in their vicinity. To lowest order in
  $\Gamma/\bar\Delta, \Gamma_{\rm d}/\bar\Delta \ll
  1$, with $\bar\Delta = \sqrt{\Omega^2 + \Delta^2}$, the $\epsilon_i$ are
  approximated by,
  \begin{equation}
  \epsilon_0 = - \Gamma_0, \qquad \epsilon_\pm = \pm i\bar\Delta - \Gamma_+,
  \end{equation}
  with effective decay rates $\Gamma_0,\Gamma_+$ given by,
  \begin{equation}
  \Gamma_0 = \Gamma \left({2 \Delta^2 + \Omega^2 \over \bar\Delta^2}\right)
  \left( 1 + {\Gamma_d \over \Gamma} \left({\Omega^2 \over 2 \Delta^2 +
  \Omega^2 }\right) \right)
  \end{equation}
and
\begin{equation}
    \Gamma_+
  = \Gamma \left({2 \Delta^2 + 3\Omega^2 \over 2\bar\Delta^2}
  \right) \left( 1 + {\Gamma_d \over \Gamma} \left({2 \Delta^2 +
\Omega^2 \over 2 \Delta^2 +
  3 \Omega^2 }\right) \right)\label{GammaPlus}.
  \end{equation}
  Thus in this limit there are three well-resolved peaks at $\omega = 0, \pm
  \bar\Delta$. Cooling is optimized for $\bar\Delta = \omega_{\rm m}$ and close to
the two
  peaks at $\pm\bar{\Delta}$ the
  real part of the spectrum can be written in the form (\ref{Eq:alphas})
with $\alpha_\pm$ generally given by,
  \begin{equation}
  \alpha_\pm = {\rm Re} \left\{ { - h(\pm \bar\Delta ) \over [\pm
  i\bar\Delta-\Gamma_0][\pm 2i \bar\Delta - \Gamma_+ ]} \right\},
  \end{equation}
which to lowest order in $\Gamma,\Gamma_{\rm d} \ll \bar\Delta$ are described
    as in (\ref{Eq:alphas}).
Note that the amplitudes of the peaks are given by $\alpha_\pm/\Gamma_+$. The analytical
expression for the real part of the spectrum will be evaluated in section
\ref{Discussion} with focus on the limit $\Gamma_{\rm d}/\Gamma\to 0$ where we can
neglect the contribution of the pure dephasing to the qubit dynamics.
For finite pure qubit dephasing the cooling rate $W$ (\ref{Eq:WForOpt}) generalizes to
\begin{equation}
\fl
W = 2\gamma_{\rm m} \left[ 1 + \beta  \tilde{f}(\varphi) \right], \qquad
\tilde{f}(\varphi) = \frac{4 \sin^2\varphi \sqrt{1 - \sin^2\varphi}}{\left(4 - \sin^4
\varphi\right) \ + \ (\Gamma_{\rm d}/\Gamma)\left(2-\sin^2\varphi\right)},
\label{Eq:WForOptApp}
\end{equation}
with
$\sin\varphi=\Omega/\bar{\Delta}$. The corresponding finite occupation number
$\langle n \rangle_{\rm f}$ (\ref{Eq:NF}) reads
\begin{equation}
\fl
\langle n\rangle_{\rm f} = \frac{N_{\rm m} \left[w_+(\varphi) \ + \ (\Gamma_{\rm
d}/\Gamma)w_-(\varphi)\right] + \beta \alpha_-}{\left[w_+(\varphi) \  + \ (\Gamma_{\rm
d}/\Gamma)w_-(\varphi)\right] + \beta (\alpha_+-\alpha_-)} , \qquad w_{\pm}(\varphi) =
{2 \pm \sin^2\varphi \over 2}, \label{Eq:NFApp}
\end{equation}
with $\alpha_{\pm}$ given in (\ref{Eq:alphas}).



\section*{References}

\begin{thebibliography}{999}

\bibitem{Arcizet2006Nature}
Arcizet O, Cohadon P F, Briant T, Pinard M and Heidmann A 2006 {\it Nature} {\bf 444} 71

\bibitem{Arcizet2006PRL}
Arcizet O, Cohadon P F, Briant T, Pinard M, Heidmann A, Mackowski J M, Michel C, Pinard L
Francais O and Rousseau L 2006 {\it Phys. Rev. Lett.}  {\bf 97} 133601

\bibitem{Gigan2006}
Gigan S, B\"ohm H R, Paternostro M, Blaser F, Langer G, Hertzberg J B, Schwab K C,
Ba\"uerle D, Aspelmeyer M and Zeilinger A 2006 {\it Nature} {\bf 444} 67

\bibitem{HoehbergerMetzger2004}
H\"ohberger-Metzger C and Karrai K 2004 {\it Nature} {\bf 432} 1002

\bibitem{Schliesser2006}
Schliesser A, Del Haye P, Nooshi N, Vahala K J and Kippenberg T J 2006 {\it Phys. Rev.
Lett.} {\bf 97} 243905

\bibitem{Kleckner2006}
Kleckner D and Bouwmeester D 2006 {\it Nature} {\bf 444} 75

\bibitem{WilsonRae2004}
Wilson-Rae I, Zoller P and Imamoglu A 2004 {\it Phys. Rev. Lett.} {\bf 92} 075507

\bibitem{Marquardt2007}
Marquardt F, Chen J P, Clerk A A and Girvin S M 2007 {\it Phys. Rev. Lett.} {\bf 99}
093902

\bibitem{Genes2008}
Genes C 2008 {\it Phys. Rev. A} {\bf 77} 033804

\bibitem{WilsonRae2007}
Wilson-Rae I, Nooshi N, Zwerger W and Kippenberg T J 2007 {\it Phys. Rev. Lett.} {\bf 99}
093901

\bibitem{Martin2004}
Martin I, Shnirman A, Tian L and Zoller P 2004 {\it Phys. Rev. B} {\bf 69} 125339

\bibitem{Zhang2005}
Zhang P, Wang Y D and Sun C P 2005 {\it Phys. Rev. Lett.} {\bf 95} 097204


\bibitem{Wang2007}
Wang Y D, Semba K and Yamaguchi H 2008 {\it New J. Phys.} {\bf 10} 043015

\bibitem{Schwab2005}
Schwab K C and Roukes M L 2005 {\it Physics Today} {\bf 58} 36

\bibitem{Marshall2003}
Marshall W, Simon C, Penrose R and Bouwmeester D 2003 {\it Phys. Rev. Lett.} {\bf 91}
130401

\bibitem{Mancini2003}
Mancini S, Vitali D and Tombesi P 2003 {\it Phys. Rev. Lett.} {\bf 90} 137901

\bibitem{Vitali2007}
Vitali D, Gigan S, Ferreira A, B\"ohm H R, Tombesi P, Guerreiro A, Vedral V, Zeilinger A
and Aspelmeyer M 2007 {\it Phys. Rev. Lett.} {\bf 98} 030405

\bibitem{Rabl2004}
Rabl P, Shnirman A and Zoller P 2004 {\it Phys. Rev. B} {\bf 70} 205304

\bibitem{Geller2005}
Geller M and Cleland A N 2005 {\it Phys. Rev. A} {\bf 71} 032311

\bibitem{Armour2002}
Armour A D, Blencowe M P and Schwab K C 2002 {\it Phys. Rev. Lett.} {\bf 88} 148301

\bibitem{Tian2005}
Tian L 2005 {\it Phys. Rev. B} {\bf 72} 195411

\bibitem{Siewert2005}
{Siewert J, Brandes T and Falci G 2005 {\it Preprint} arXiv:cond-mat/0509735}


\bibitem{LaHaye2004}
LaHaye M D, Buu O, Camarota B and Schwab K C 2004 {\it Science} {\bf 304} 74

\bibitem{Cohadon1999}
Cohadon P F, Heidmann A and Pinard M 1999 {\it Phys. Rev. Lett.} {\bf 83} 3174

\bibitem{Naik2006}
Naik A, Buu O, LaHaye M D, Armour A D, Clerk A A, Blencowe M P and Schwab, K C 2006 {\it
Nature} {\bf 443} 193

\bibitem{Blencowe2005}
Blencowe M P, Imbers J and Armour A D 2005 {\it New J. Phys.} {\bf 7} 236

\bibitem{Clerk2005}
Clerk A A and Bennett S 2005 {\it New J. Phys.} {\bf 7} 238

\bibitem{Knobel2003}
Knobel R G and Cleland A N 2003 {\it Nature} {\bf 424} 291

 \bibitem{Teufel2008}
 {Teufel J D, Regal C A and Lehnert K W 2008 {\it Preprint} arXiv:0803.4007}


\bibitem{Blencowe2007}
Blencowe M P and Buks E 2007 {\it Phys. Rev. B} {\bf 76} 014511

\bibitem{Wineland2006}
{Wineland D J, Britton J, Epstein R J, Leibfried D, Blakestad R B, Brown K,
Jost J D, Langer C, Ozeri R, Seidelin S and Wesenberg J 2006 {\it Preprint}
arXiv:quant-ph/0606180}

\bibitem{Vion2002}
Vion D, Aassime A, Cottet A, Joyez P, Pothier H, Urbina C, Esteve D, Devoret M H 2002
{\it Science} {\bf 296} 886

\bibitem{Stenholm1986}
Stenholm S, 1986 {\it Rev. Mod. Phys.} {\bf 58} 699

\bibitem{Cirac1992}
Cirac J I, Blatt R, Zoller P and Phillips W D 1992 {\it Phys. Rev. A} {\bf 46} 2668

\bibitem{Cleland2003}
Cleland A N 2003 {\it Foundations of Nanomechanics} (Berlin: Springer)

\bibitem{MathMethods}
Barnett S M and Radmore P M 1997 {\it Methods in Theoretical Quantum Optics} (Oxford:
Oxford University Press)

\bibitem{Lafarge1993}
 Lafarge P, Loyez P, Esteve D, Urbina C and Devoret M H 1993 {\it Nature} {\bf 365} 422

\bibitem{Shnirman1997}
Shnirman A, Sch\"on G and Hermon Z 1997 {\it Phys Rev Lett} {\bf 79} 2371

\bibitem{Makhlin2001}
Makhlin Y, Sch\"on G and Shnirman A 2001 {\it Rev. Mod. Phys.} {\bf 73} 357

\bibitem{Ithier2005}
Ithier G, Collin E, Joyez P, Meeson P J, Vion D, Esteve D, Chiarello F, Shnirman A,
Makhlin Y, Schriefl J and Sch\"on G 2005 {\it Phys. Rev. B} {\bf 72} 134519

\bibitem{Wallquist2005}
Wallquist M, Lantz J, Shumeiko V S and Wendin G 2005 {\it New J. Phys.} {\bf 7} 178

\bibitem{WallquistThesis}
Wallquist M 2006 {\it Controllable Coupling of Superconducting Qubits and Implementation
of Quantum Gate Protocols} (Chalmers University of Technology: PhD thesis)

\bibitem{Duty2004}
Duty T, Gunnarsson D, Bladh K and Delsing P 2004 {\it Phys. Rev. B} {\bf 69} 140503 (R)

\bibitem{LandauLifshitz}
Landau L D, Lifschitz E M 1974 {\it Elektrodynamik der Kontinua} (Berlin: Akademie
Verlag)

\bibitem{Cleland1996}
Cleland A N and Roukes M L 1996 {\it Appl. Phys. Lett.} {\bf 69} 2653

\bibitem{WallsMilburn}
Walls D F and Milburn G J 1994 {\it Quantum Optics} (Berlin: Springer)

\bibitem{Hauss2008}
Hauss J, Fedorov A, Hutter C, Shnirman A and Sch\"on G 2008 {\it Phys. Rev.
Lett.} {\bf 100} 037003


\bibitem{Devoret1997}
Devoret M 1997 {\it Quantum fluctuations in electrical circuits} (Les Houches LXIII,
1995) (Amsterdam: Elsevier)

\bibitem{Tinkham1996} Tinkham M 1996 {\it Introduction to superconductivity} (New York:
McGraw-Hill)

\bibitem{Goldstein}
Goldstein H, Poole Jr C P and Safko J L  2002 {\it Classical Mechanics} (San Francisco:
Addison-Wesley)





\end{thebibliography}
\end{document}